\documentclass[unpublished,a4paper,onecolumn,superscriptaddress,11pt]{revtex4-1}
\usepackage{amsmath}
\usepackage{amsfonts}
\usepackage{amssymb}
\usepackage{bm}
\usepackage{MnSymbol,wasysym}
\usepackage[pdftex]{graphicx}
\usepackage{epstopdf}
\usepackage{etoolbox}
\usepackage{times}
\usepackage[letterpaper,margin=1in]{geometry}
\usepackage[utf8]{inputenc}
\usepackage{color}
\usepackage[usenames,dvipsnames]{xcolor}
\usepackage{epstopdf}
\usepackage{units}
\usepackage[normalem]{ulem}
\usepackage{enumerate}
\usepackage[caption=false]{subfig}
\usepackage{longtable}
\usepackage{float}
\usepackage{braket}
\usepackage{hyperref}
\usepackage{silence}
\usepackage[utf8]{inputenc}
\definecolor{Red}{rgb}{1,0,0}
\usepackage[section]{placeins}

\begin{document}
\title{Measuring the time atoms spend in the excited state due to a photon they don't absorb}

\author{Josiah Sinclair}
\affiliation{Department of Physics, and Centre for Quantum Information and Quantum Control,  University of Toronto, 60 St. George Street, Toronto, Ontario, Canada M5S 1A7}
\author{Daniela Angulo}
\affiliation{Department of Physics, and Centre for Quantum Information and Quantum Control,  University of Toronto, 60 St. George Street, Toronto, Ontario, Canada M5S 1A7}
\author{Kyle Thompson}
\affiliation{Department of Physics, and Centre for Quantum Information and Quantum Control,  University of Toronto, 60 St. George Street, Toronto, Ontario, Canada M5S 1A7}
\author{Kent Bonsma-Fisher}
\affiliation{Department of Physics, and Centre for Quantum Information and Quantum Control,  University of Toronto, 60 St. George Street, Toronto, Ontario, Canada M5S 1A7}
\affiliation{National Research Council of Canada, 100 Sussex Drive, Ottawa, Ontario, Canada K1A 0R6}
\author{Aharon Brodutch}
\affiliation{Department of Physics, and Centre for Quantum Information and Quantum Control,  University of Toronto, 60 St. George Street, Toronto, Ontario, Canada M5S 1A7}
\author{Aephraim M.  Steinberg}
\affiliation{Department of Physics, and Centre for Quantum Information and Quantum Control,  University of Toronto, 60 St. George Street, Toronto, Ontario, Canada M5S 1A7}
\affiliation{Canadian Institute For Advanced Research, 180 Dundas St. W., Toronto, Ontario, Canada, M5G 1Z8}

\date{\today}

\begin{abstract}

When a resonant photon traverses a sample of absorbing atoms, how much time do atoms spend in the excited state?  Does the answer depend on whether the photon is ultimately absorbed or transmitted? In particular, if it is {\it not} absorbed, does it cause atoms to spend any time in the excited state, and if so, how much? In an experiment with ultra-cold Rubidium atoms, we simultaneously record whether atoms are excited by incident (``signal'') photons and whether those photons are transmitted. We measure the time spent by atoms in the excited state by using a separate, off-resonant ``probe" laser to monitor the index of refraction of the sample -- that is, we measure the nonlinear phase shift written by a 
signal pulse on this probe beam -- and use direct detection to isolate the effect of single transmitted photons. For short pulses ($10$ ns, to be compared to the $26$ ns atomic lifetime) and an optically thick medium (peak OD = 4, leading to 60\% absorption given our broad bandwidth), we find that the average time atoms spend in the excited state due to one transmitted photon is not zero, but rather $(77 \pm 16)\%$ of  the time the average incident photon causes them to spend in the excited state. We attribute this observation of ``excitation without loss'' to coherent forward emission, which can arise when the instantaneous Rabi frequency (pulse envelope) picks up a phase flip -- this happens naturally when a broadband pulse propagates through an optically thick medium with frequency-dependent absorption \cite{Crisp1970}. These results unambiguously reveal the complex history of photons as they propagate through an absorbing medium and illustrate the power of  utilizing post-selection to experimentally investigate the past behaviour of observed quantum systems. 

\end{abstract}

\maketitle

When a resonant photon passes through a sample of atoms, the photon can either be absorbed (``lost'') or transmitted. 
``Loss'' in fact involves two steps: excitation of an atom by a photon, and subsequent decay of the atom (whether via spontaneous emission to a side mode, non-radiative decay, or other).
In free space and in the absence of stimulated emission, re\"emission into the forward mode might seem such an unlikely event as to suggest that atomic excitation is {\it almost always} accompanied by loss. In other words, in the cases where a photon is successfully transmitted through the absorbing medium, should we conclude it was transmitted because it was lucky enough never to excite an atom? On the other hand, one could take the point of view that photons are simply an electromagnetic field which polarizes atoms -- irrespective of whether they are ultimately to be transmitted or lost -- thus creating a small excitation probability wherever they are present in the medium. In this latter, ``egalitarian,'' picture,
lost and transmitted photons still behave differently, because the former are on average scattered within the first optical decay length, while the latter have the opportunity to polarize the entire sample. Does this mean that transmitted photons 
cause atoms to cumulatively spend {\it more} time in the excited state than do absorbed ones?
Questions about the past behaviour of a quantum particle, or about trajectories in quantum mechanics, have a famously controversial history with important implications for the foundations and interpretation of quantum physics \cite{Bohm1952, ABL1964, wheeler_zurek_1983, Griffiths1984, AAV1988,Leggett1989, gellmann2018quantum, Mir_2007, Kocsis1170, Lundeen2009, Vaidman2013}. Renewed interest in these questions in the context of open quantum systems has also motivated new theoretical approaches \cite{Gammelmark2014, Guevara2018,Tsang2009} and experiments \cite{Rybarczyk2015, HacohenGourgy2016, Minev2019}.

To study the degree of excitation produced in a cloud of cold $^{85}$Rb atoms by an incident signal pulse, we use a nonlinear optical effect known as a resonant Kerr nonlinearity, widely studied for the ``cross-phase shifts'' (XPS) it can generate, with applications for instance to quantum logic gates \cite{Kimble1995, Schmidt96,Venkataraman2013, Feizpour2015a, Tiarkse1600036, Sinclair2019}.  A continuous-wave, off-resonant, probe beam overlapped with -- but counter-propagating with respect to -- the signal pulse acquires a phase that depends on the population inversion, due to the different indices of refraction of the ground and excited states. The time-integral of the phase shift relative to its value when no atoms are excited is proportional to the ``excitation time,'' by which we mean the expectation value of the total time spent cumulatively by all atoms in the excited state: the time-integral of the mean occupation number in the excited state. In principle, exciting a single atom would change the phase of a resonant laser beam focused down to one atomic cross-section by an amount on the order of a radian. In our experiment, the probe is focused to approximately $10^{4}$ cross-sections, and acquires a phase of a few tens of $\mu$rad per excited atom.  This phase shift, which lasts on the order of a 26-ns atomic lifetime, is currently far too small for us to observe on a single shot.

To resolve such a small effect, one could imagine carrying out this measurement on a sequence of single incident photons, binning the measurements of excitation time according to whether the photon was transmitted or lost, and averaging \footnote{this is clearly closely related to the weak measurements of Aharonov, Albert and Vaidman, as it involves a pre- and post-selection, and a weak coupling to a probe system; for other examples of pre- and post-selected systems including from Cavity QED see \cite{Foster2000,Smith2002, Wiseman2002, Duan2020})}. Although there exist sources of single, narrow-band photons \cite{Thompson74,Xing2013,Loredo2016,JianWeiPan2016,ornelashuerta2020}, they are currently limited to kHz rates. Instead, we took advantage of the faster rates (approximately $2$ MHz) that can be achieved  using pulsed coherent states by relying on a novel quantum effect discovered by our group several years ago \cite{Feizpour2015a}. In this effect, the mean number of photons one should infer were present in an interaction region illuminated by a coherent state $|\alpha\rangle$ increases by one (aside from minor corrections due to saturation and dark counts) whenever a photon is detected.
(In the limit of small incident mean photon number $|\alpha|^2 \ll 1$, it is clear that each detection event revises one's estimate to {\it at least} $1$; more generally, when a coherent state is split into two modes -- e.g., the ``detected'' and ``undetected'' modes for an inefficient photon-counter -- their states remain separable, meaning that a measurement collapses the ``detected'' mode to have $0$ or $1$ photons while leaving the mean number in the ``undetected'' mode unchanged.) We first employed this effect to isolate the cross-phase shift due to a post-selected single photon \cite{Feizpour2015a}, and shortly after, demonstrated Weak Value Amplification of the cross-phase shift due to a post-selected single photon \cite{Hallaji2017}. In the current experiment, when a photon is detected after the medium, the inferred mean number of \textit{transmitted} photons is increased from $|t\alpha|^2 \rightarrow |t\alpha|^2 + 1$, where $|t|^2$ is the probability for a photon to be transmitted and $|\alpha|^2$ is the mean number of resonant photons entering the medium. By binning our measurements of excitation time in terms of whether a photon is detected or not, we can isolate the effect of a transmitted photon.

\begin{figure}[t]
 \centering
\includegraphics[width=1\textwidth]{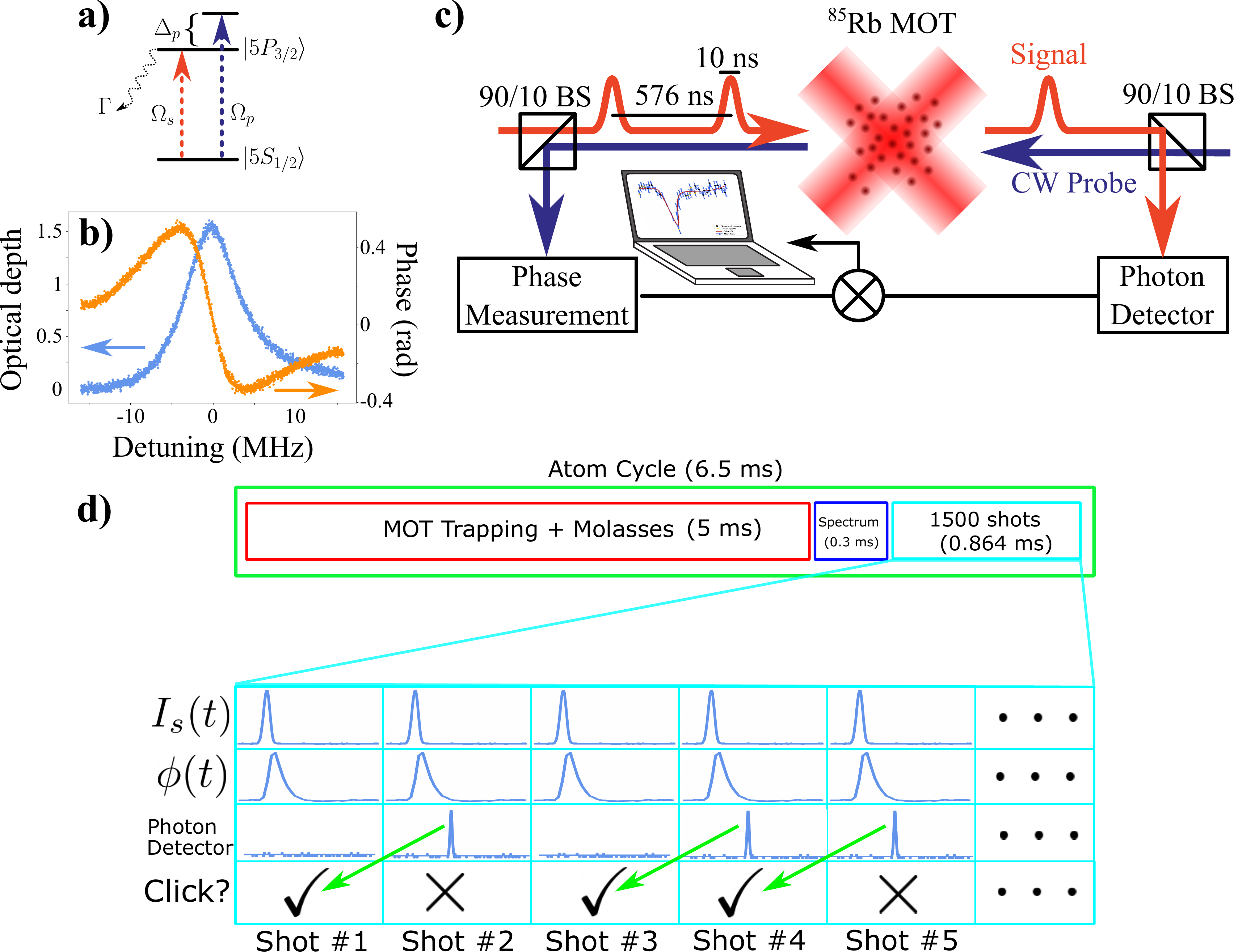}
                \caption{(a) \textbf{Level scheme used} in the experiment;  $\Omega_{p(s)}$ is the probe (signal) Rabi frequency, $\Delta_p$ is the probe detuning, and $\Gamma$ is the spontaneous decay rate of the excited state. (b) \textbf{Measured optical depth and phase} experienced by the probe as a function of $\Delta_p$. (c) \textbf{Simplified schematic of the  experimental apparatus} consisting of signal and probe beams overlapped using 90-10 beamsplitters, the Rubidium MOT, photon detection, and phase measurement. (d) \textbf{Experimental timing sequence} used in the experiment. Green arrows indicate photon detection being re-associated with the correct shot. $I_s(t)$ is the average signal intensity measured with a fast APD, $\phi(t)$ is the measured phase acquired by the probe in a shot, and ``Click?" indicates whether a photon was detected or not. 
                \label{Apparatus}}
                
\end{figure}

For the experiment (see Fig.~\ref{Apparatus}), we use an ultra-cold cloud of $^{85}\textrm{Rb}$ in a magneto-optical trap (MOT) and a pair of counter-propagating, overlapping beams focused to about $25\,  \mu$m inside the atom cloud. We set the signal on resonance with the $F=3$ to $F'=4$ transition of the 780 nm D2 line and detune the probe by $\pm (3 - 10)$ MHz from the same transition to monitor the excited-state population (Fig.~\ref{Apparatus}a shows the level scheme used). The beams are spatially overlapped and they are separated using 90-10 beamsplitters. After passing through the MOT, the probe is collected in multimode fibre. We measure the phase using frequency domain interferometry, which involves beating the probe against a copropagating reference shifted off-resonance by $100$ MHz, and hence nearly unaffected by the atoms.

Each cycle of the experimental timing sequence (Fig. \ref{Apparatus}d) consists of approximately $5$ ms of MOT trapping, molasses and free expansion, and a $1.152$ ms ``measurement" stage.
During the first $288$ ns of each measurement stage, the probe phase and transmission are measured while the frequency of the probe (and reference) is scanned across resonance to extract the optical depth (OD) and phase as a function of detuning (see Fig. \ref{Apparatus}b for a measurement of the probe transmission spectrum taken at reduced OD). For the remaining $864 \: \mu$s, the probe is left on continuously while signal pulses pass through the medium. Signal pulses have a Gaussian temporal profile with an rms width of $10$ ns and are separated by $576$ ns (such that the $864 \, \mu$s acquisition window contains 1500 pulses). Each $576$ ns window is referred to as a ``shot,'' and contains 36 phase measurements  taken at $16$ ns intervals, enabling us to separate the phase shift due to the signal pulse from slower variations in the index of refraction.

\begin{figure}[t]
 \centering
\includegraphics[width=1\textwidth]{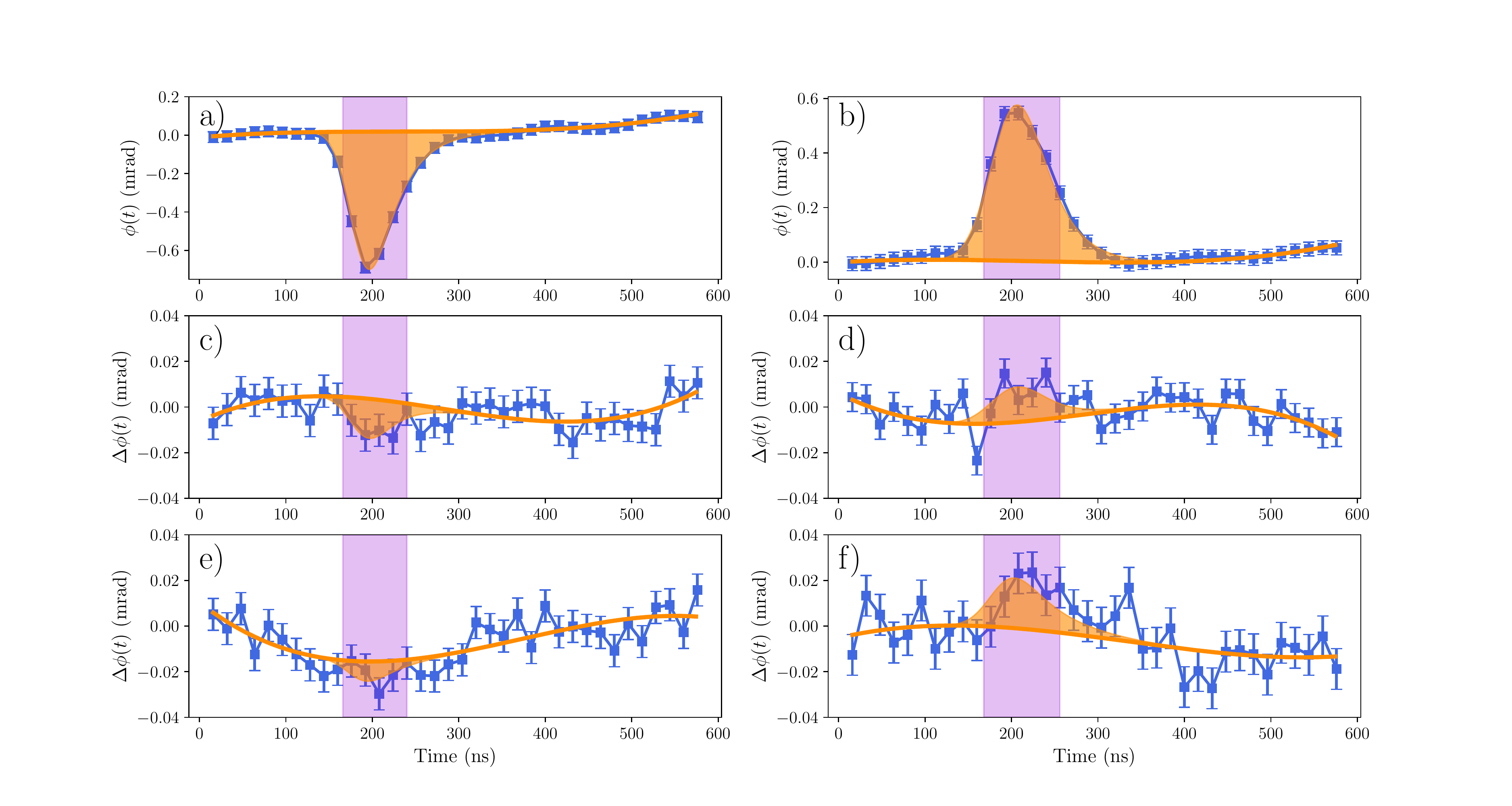}
                \caption{\textbf{Probe phase shift (a, b) and probe phase shift differences (c, d, e, f) versus time}, averaged over approximately six billion shots, in the presence of a 10-ns signal pulse.  Probe phase shifts are recorded for pulses, containing 34 photons on average, incident at approximately $t=200$ ns for probe detunings of (a) $-5.6$ MHz and (b) $+4.7$ MHz.  The purple rectangle indicates the window of time wherein the cross-phase shift occurs.
                The shaded gold curve is a fit to a cubic polynomial (to model the background drift) plus a peaked function. The solid gold line shows only the cubic fit to background.
                Plots (c) through (f) show $\Delta \phi(t)$ for $34$-photon pulses in (c) and (d) and $134$-photon pulses in (e and f).  As before, the probe detuning is negative on the left (c and e) and positive on the right (d and f).  
                Here, the shaded gold region indicates a fit to a cubic polynomial plus a function with the shape of the peaks fitted to the relevant average XPS (e.g. for c, the XPS in (a) was used, while the XPS used for (e and f) is not shown), but with an adjustable amplitude.  As before, the solid gold line shows only the cubic background fit. The phase is sampled every $16$ ns and a $25$ MHz measurement bandwidth is used. The phase noise on a single sample is approximately $100-200$ mrad and is determined by the sampling rate, measurement bandwidth, and the probe power, which was set to approximately $5$ nW in order to be much less than saturation (about $50$ nW). Error bars shown are the standard error of the mean.   
                \label{rawdata}}
                
\end{figure}

The average phase acquired by the probe within a shot ($\phi(t)$)  in the presence of a 34-photon signal pulse is plotted in Fig. \ref{rawdata}a for a red-detuned probe ($-5.6$ MHz, negative cross-phase shift) and in \ref{rawdata}b for a blue-detuned probe ($+4.7$ MHz, positive cross-phase shift). In both \ref{rawdata}a and \ref{rawdata}b, the 576-ns trace has been averaged approximately six billion times and the additional phase shift due to the 34-photon signal pulse is clearly visible, with a peak of $\pm 700 \: \mu$rad. The pulse enters at about $175$ ns into the shot and $\phi(t)$, which is a measure of the excited-state population, grows quickly as the signal pulse enters the medium (with a slight delay due to the 25 MHz bandwidth of our data acquisition), peaks around $200$ ns, and decays exponentially over a slightly longer time-scale set by the atomic lifetime ($26.5$ ns) after the signal pulse has exited the medium. The peak cross-phase shift per photon, $\phi_0$, is found by dividing the peak excursion of $\phi(t)$ from the background phase by the average number of photons in the signal pulse. $\phi_0$ is $-20.0 \pm 0.4 \: \mu$rad/photon and $+16.4 \pm 0.4 \: \mu$rad/photon in Figures~\ref{rawdata}a and \ref{rawdata}b, respectively.

Once it exits the medium, the transmitted portion of the signal is collected in a single-mode fibre and detected with a single-photon counting module (SPCM) (see Fig. \ref{Apparatus}c) with a detector efficiency of about $70\%$. 
The total path efficiency is about 20\%, but to reduce spurious counts from probe light leaking into the fibre, the signal collection efficiency is further reduced using a variable neutral-density filter so that background counts are suppressed to about $1\%$ per shot (about 20,000 cps).  The number of input signal photons is adjusted to maintain signal detection probability between $20-30\%$ (typical photon numbers were $34$ and $134$). In order to prevent electrical cross-talk, signal detection is optically delayed by about $600$ ns. Detection events are digitized in parallel with the phase, and associated (re-aligned) with the correct shot during analysis (see Fig. \ref{Apparatus}d). 

We analyze our data by binning all shots into ``shots with a click" or ``shots without a click" depending on whether or not the SPCM detected a photon associated with that shot. $\phi(t)$ for all the shots ``with a click" is then averaged -- this is called $\phi_C(t)$ -- and subtracted from the average of all the shots ``with no click" -- this is called $\phi_{NC}(t)$. This difference, $\Delta \phi(t) = \phi_C(t) - \phi_{NC}(t)$,
isolates the effect of a transmitted single photon on the probe's phase (which is to say, on the excited-state population). But $\Delta \phi(t)$ is not sensitive only to the effect of a transmitted photon. \textit{Any} correlation between the phase of the probe and the click probability (i.e., the signal transmission) will show up in $\Delta \phi(t)$ -- even small spurious correlations, which have the potential to reduce, swamp, or even masquerade as the effect of a single transmitted photon.

Spurious correlations can arise from two sorts of effects: (1) fluctuations which affect both the {\it linear} phase shift experienced by the probe and the transmission of the signal, and which occur on short enough timescales that our background subtraction fails to eliminate them; and (2) fluctuations which affect the {\it magnitude} of the {\it nonlinear} cross-phase shift and the transmitted signal power (whether they are fast or slow).

Fluctuations of the first kind can have a huge effect, even if they are quite small, because the linear phase picked up by the probe passing through the atom sample is on the order of a radian -- 1000 times larger than the XPS imparted by a signal pulse containing approximately $100$ photons. In our experiment, we observed features in $\Delta \phi(t)$, which we attributed to correlations of this kind because of their particular dependence on signal detuning (anti-symmetric around resonance) and probe detuning (symmetric around resonance). We additionally observed that the time dependence of these features differed from that of the nonlinear cross-phase shift, displaying damped oscillatory behavior with a period of approximately $500$ ns and an amplitude that diminished with decreasing pulse length. (If there is a background fluctuation on a timescale longer than the pulse length, no correlations build up between transmission and linear phase.) We therefore carried out the experiment with short pulses so that any effects on $\Delta \phi(t)$ due to spurious correlations which remain due to the signal not being perfectly on resonance (where their effect disappears) would be both suppressed compared to, and distinguishable from, the sharply-peaked cross-phase shift due to a transmitted photon. In Fig. \ref{rawdata}c-f, the residual effects of this kind of spurious correlation appear as a slowly varying background in $\Delta \phi(t)$. We reject this slowly-varying background by fitting all $36$ phase difference measurements to a cubic function (gold line), plus a peaked function whose amplitude is left as a free parameter $\phi_T$ but whose shape is constrained to match the shape of $\phi(t)$ (gold shading).
Figures \ref{rawdata}c and \ref{rawdata}e show $\Delta \phi(t)$ for signal pulses with  $|\alpha|^2 = 34$  and $|\alpha|^2 = 134$  photons respectively, both of which were carried out with a negative probe detuning (and therefore a negative $\phi_0$). Fig. \ref{rawdata}d and \ref{rawdata}f also depict $\Delta \phi(t)$ taken with $|\alpha|^2 = 34$  and $|\alpha|^2 = 134$ photons, but were taken with a positive probe detuning (and therefore a positive $\phi_0$). Each plot is the result of averaging together roughly 6 billion shots (and as many signal pulses), requiring a total of $900$ billion phase measurements and over one hundred hours of data acquisition.

Fluctuations of the second kind can contribute even if they are very slow, as they lead to the XPS being correlated with detection probability, and are therefore not eliminated by our background-subtraction procedure. Two such mechanisms include (1) signal intensity fluctuations beyond those inherent to a coherent state, and (2) variations in atom number (OD). If some signal pulses are brighter than others, they will naturally lead to higher detector click rates \textit{and} to larger nonlinear phase shifts. Since this is an increase in the phase shift written by all of the $|\alpha|^2$ photons in the pulse, its effect can easily exceed the signal of interest due to an individual transmitted photon. Similarly, if the number of atoms fluctuates, periods of high optical depth will be correlated with both low signal transmission and high nonlinear phase shifts. Such correlations, unlike those of the first kind, cannot be distinguished from the effect of a transmitted photon through their dependence on detuning, or their time dependence, both of which are identical to that of the XPS. Instead, the degree of noise must be independently characterized.  During the experiment, we continuously monitored the signal optical depth and observed optical depth fluctuations of less than $5\%$. To assess the total effect of proportional noise in our experiment, we measure $\Delta \phi(t)$ for signal pulses with $588$, $898$, $1527$ and $3040$ photons, and from these measurements estimate the total proportional noise due to the combined effects of OD variation and signal intensity fluctuations beyond that of a coherent state to be less than or equal to about $3\%$, substantially less than Poisson fluctuations for coherent states with average photon number of $34$ or $134$.

\begin{figure*}[t]
 \centering
\includegraphics[width=1\textwidth]{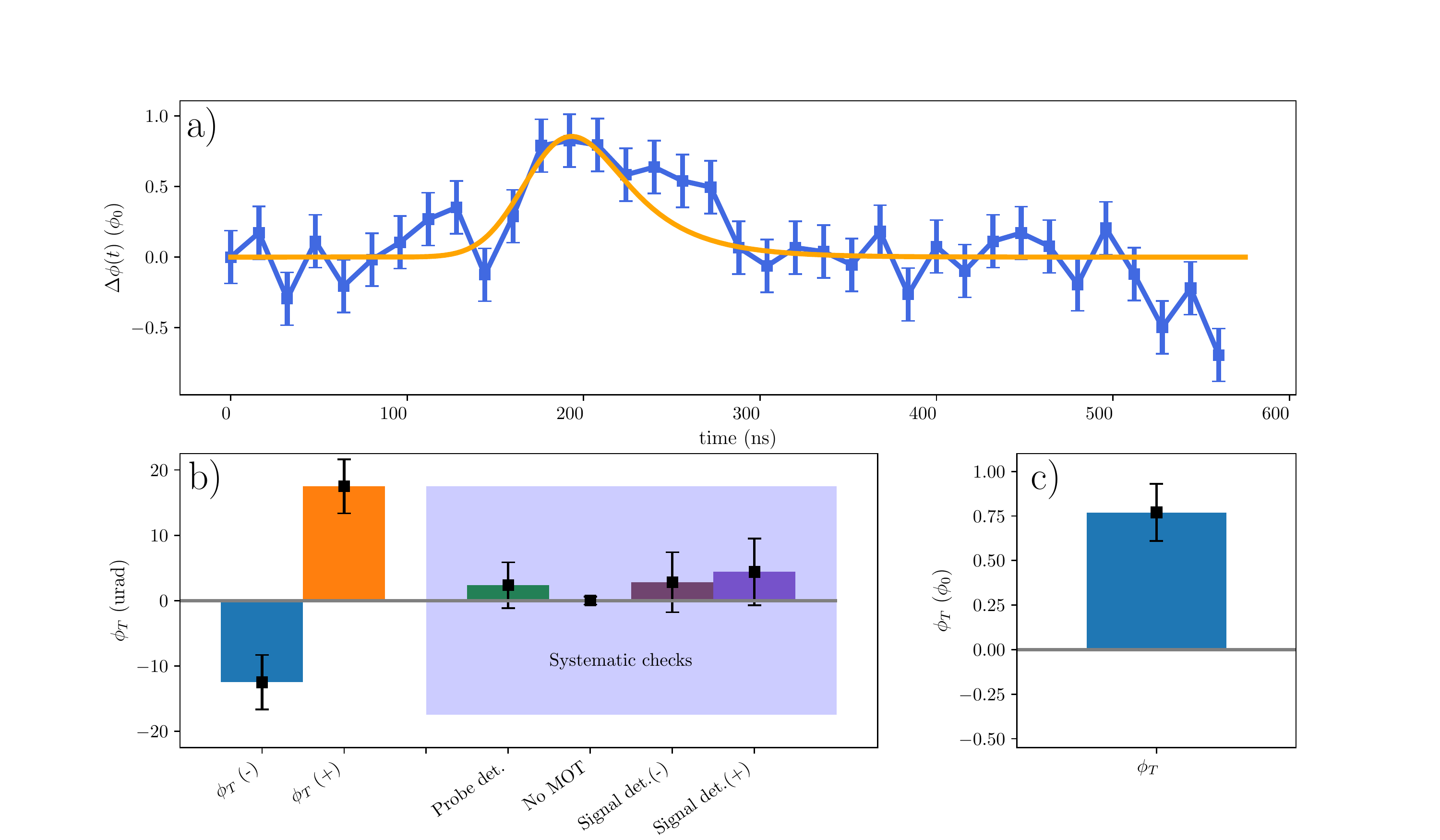}
                \caption{(a) \textbf{A weighted average} of ${\Delta} {\phi(t)}$ with $34$ and $134$ photons, and with positive and negative probe detunings, in units of $\phi_0$. The gold line is the shape of the cross-phase shift taken from the fits to $\phi(t)$, and is meant to guide the eye. (b) \textbf{The phase shift due to a transmitted photon} ($\phi_T$) for $-5.6$ MHz and $+4.7$ MHz is shown. In the shaded region, the null results of checks for systematic effects are shown: $\phi_T$ with the probe detuned far from resonance, with the MOT turned off, and with the signal detuned far from resonance (with the probe detuning set to be above (+) and below (-) resonance). (c) \textbf{Our best estimate of the relative phase shift due to a transmitted photon} ($\phi_T/\phi_0$), found by combining measurements from (b) at negative and positive detunings. 
                \label{ExperimentalResults1}}
                
\end{figure*}

In Fig. \ref{ExperimentalResults1}a all of the data from Figures \ref{rawdata} c-f are combined and normalized to the relevant $\phi_0$ for each curve.
Below, Fig. \ref{ExperimentalResults1}b shows the peak amplitude $\phi_T$, after correction for the effects of proportional noise, for both the positive- and negative-detuning cases. The shaded region shows the null results of checks for systematic effects which might masquerade as that of a single photon: data taken with the probe detuned far from resonance; with the MOT turned off; and with the signal detuned far from resonance (with the probe detuned to the red (-) and to the blue (+)). In Fig. \ref{ExperimentalResults1}c, measurements at positive and negative detunings are combined to produce our best estimate of the XPS due to a transmitted photon as a fraction of the XPS due to an average incident photon: we find $\phi_T/\phi_0 = 0.77 \pm 0.16$. The unequivocal implication of this result is that even photons which are ultimately transmitted spend a significant portion of time as atomic excitations.

As unexpected as ``excitation without loss" might be, even less intuitive is the implication that lost photons are scattered long before the atom's spontaneous lifetime has elapsed. This conclusion follows directly from the fact that measurements of $\phi_0$ and $\phi_T$ are reflective of the time atoms spend in the excited state ($\tau_0$ and $\tau_T$) and the fact that spontaneous emission occurs at a constant rate $\Gamma=1/\tau_{\rm sp}$, making a photon's probability of being lost proportional to the total time it causes atoms to spend in the excited state: $P_L = \Gamma\tau_0 = \tau_0/\tau_{\rm sp}$. In other words, the total time spent in the excited state due to an average incident photon is equal to the product of the loss probability and the atomic lifetime $\tau_{\rm sp}$.   But of course $\tau_0$ must also be equal to the average of the excitation times for lost and transmitted photons, $\tau_L$ and $\tau_T$, weighted by the loss and transmission probabilities.  Thus, any non-zero (and positive) $\tau_T$ immediately implies that $\tau_L < \tau_{\rm sp}$, which can only occur if atoms spend on average less than one atomic lifetime in the excited state before spontaneously emitting.

For a broadband pulse passing through an optically thick medium, $\tau_L < \tau_{sp}$ proves to be easy to understand: when such a  pulse propagates through an absorbing medium, the envelope acquires a phase flip which causes the pulse area  --  the angle through which this pulse will rotate an atomic Bloch vector  --  to decay quickly to zero, even in regions where the pulse energy is not decaying rapidly.  (This can be understood through the pulse area theorem \cite{McCall1967a,McCall1967b,AllenEberlyMonograph}; or in terms of temporal beats created once the central part of the spectrum is absorbed; and/or as atomic emission which persists long after the passage of the initial pulse, but is $180^\circ$ out of phase with it \cite{Crisp1970}.) Regardless of the perspective one takes, the consequence for the atoms is that portions of the pulse -- even in the linear, or single-photon, regime \cite{Costanzo2016} -- coherently drive excited atoms back to the ground state. We believe that this coherent forward emission process underlies both the nonlinear phase shift
associated with signal photons that are not lost, {\it and} the {concomitant} reduction of the mean time spent in the excited state below $\tau_{\rm sp}$ (which drives the nonlinear effect of lost photons below what would be na\"ively expected).  
While such semiclassical arguments support the notion that coherent emission must be present, they are incapable of  quantifying precisely how much of it occurs, and provide only a lower bound.  Moreover, they are of course silent on questions of entanglement between the final state of the signal photon and the time spent by an atom in the excited state.

\begin{figure*}[h]
 \centering
\includegraphics[width=1\textwidth]{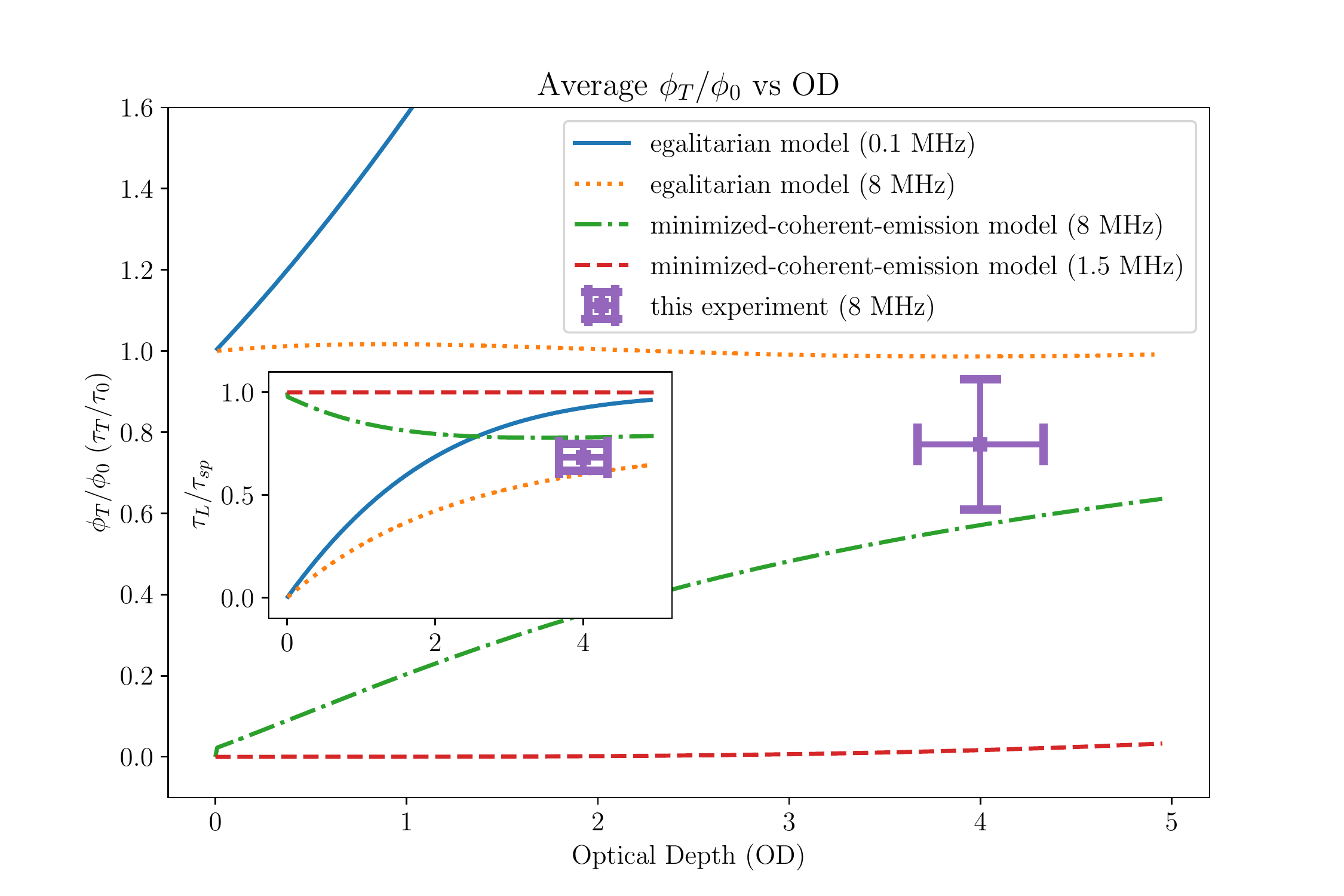}
                \caption{\textbf{A comparison of our experimental results to the predictions of various semi-classical models.}  
                  The predictions of a toy model in which coherent forward emission occurs as a result of an interference effect arising when broadband small-area pulses propagate through an optically thick medium are shown  
                  both for signal pulses of $8$ MHz bandwidth as in our experiment (green dot-dashed line), and for narrowband ($1.5$ MHz) ones (red dashed line). 
                  An ``egalitarian'' model in which photons affect the medium up until the point where they are scattered (or transmitted) is shown for the $8$ MHz  (orange dotted line) and $0.1$ MHz (blue solid line) cases. Our experimental results are also shown (purple), with the peak on-resonant OD inferred from the optical depth seen by the probe (about 4), consistent with the amount of attenuation experienced by a resonant pulse with a bandwidth of $8$ MHz. In the inset, $\tau_L/\tau_{sp}$ is plotted for all four models, along with the inferred result of our measurement. 
                \label{Theory}}
                
\end{figure*}

In Figure 4, we compare our experimental result (purple) with several simple theoretical models: in blue [solid] for narrowband pulses (orange [dotted] for broadband pulses), the egalitarian model introduced earlier, in which absorbed photons affect the medium only up until the point where they are scattered, whereas transmitted photons interact with the full length of the medium; in green [dot-dash], a model based on the minimum amount of coherent forward emission consistent with the semiclassical Bloch-state evolution for a 10-ns pulse like the one used in our experiment; and in red [dashed], for comparison, the predictions such a minimized-coherent-emission model would make for a narrowband, $50$-ns pulse. The data clearly rule out $\phi_T \propto \tau_T =0$, and are consistent with a range of values which includes both the egalitarian model and the minimized-coherent-emission prediction.  These models, however, make strikingly different predictions for $\tau_T$ in the narrow-bandwidth limit; and for $\tau_L$ in the low-OD limit  (see inset).

The ``egalitarian'' model predicts that $\tau_T \neq 0$, and hence that $\tau_L < \tau_{\rm sp}$.  Its implications, however, become extreme in the OD $\rightarrow 0$ limit.  Here $P_L \rightarrow$ OD, and $\tau_L \rightarrow \tau_T/2$ (because the average position for a photon to be scattered is halfway through the medium).  As the transmission probability approaches $100\%$, $\tau_0 \rightarrow \tau_T$; and yet we know $\tau_0 = P_L \tau_{\rm sp}$. This would imply that $\tau_T/\tau_{\rm sp} \rightarrow {\rm OD}$ and $\tau_L /\tau_{\rm sp} \rightarrow {\rm OD}/2 \rightarrow 0$.  So this model not only requires a mechanism to pull atoms out of the excited state faster than $\Gamma$, but needs the speed of this mechanism to {\it diverge} as $1/$OD, which seems impossible to justify on physical grounds.  More {natural} would be the expectation that {\it at least} in the low-OD regime, $\tau_L \rightarrow \tau_{\rm sp}$, implying $\tau_T \rightarrow 0$, as occurs for instance in the minimum-coherent-emission model we base on the solutions to the Maxwell-Bloch equations. The latter model has extreme sensitivity not only to bandwidth but also to pulse shape and duration (which is unusual for calculations of the effect of a single photon), 
but it reproduces the intuition that for low OD or narrowband pulses, $\tau_L \rightarrow \tau_{\rm sp}$.

The experiment described in this work constitutes an important milestone: the first measurement of the time atoms spend in the excited state due to photons they don't ultimately absorb. We find that this time is nonzero, and can even be on the order of the atomic lifetime. These results cry out for a fully quantized theoretical treatment as well as further experiments varying the optical depth, bandwidth, and pulse shape in order to further elucidate the strange history of transmitted photons.

\subsection{Acknowledgements}
We would like to thank Alan Migdall, Howard Wiseman, John Sipe, Amir Feizpour, Amar Vutha, and Joseph Thywissen for helpful conversations.  This work was supported by NSERC and the Fetzer Franklin Fund of the John E. Fetzer Memorial Trust; AMS is a Fellow of CIFAR.

\bibliography{CvNC}

\begin{thebibliography}{39}%
\makeatletter
\providecommand \@ifxundefined [1]{%
 \@ifx{#1\undefined}
}%
\providecommand \@ifnum [1]{%
 \ifnum #1\expandafter \@firstoftwo
 \else \expandafter \@secondoftwo
 \fi
}%
\providecommand \@ifx [1]{%
 \ifx #1\expandafter \@firstoftwo
 \else \expandafter \@secondoftwo
 \fi
}%
\providecommand \natexlab [1]{#1}%
\providecommand \enquote  [1]{``#1''}%
\providecommand \bibnamefont  [1]{#1}%
\providecommand \bibfnamefont [1]{#1}%
\providecommand \citenamefont [1]{#1}%
\providecommand \href@noop [0]{\@secondoftwo}%
\providecommand \href [0]{\begingroup \@sanitize@url \@href}%
\providecommand \@href[1]{\@@startlink{#1}\@@href}%
\providecommand \@@href[1]{\endgroup#1\@@endlink}%
\providecommand \@sanitize@url [0]{\catcode `\\12\catcode `\$12\catcode
  `\&12\catcode `\#12\catcode `\^12\catcode `\_12\catcode `\%12\relax}%
\providecommand \@@startlink[1]{}%
\providecommand \@@endlink[0]{}%
\providecommand \url  [0]{\begingroup\@sanitize@url \@url }%
\providecommand \@url [1]{\endgroup\@href {#1}{\urlprefix }}%
\providecommand \urlprefix  [0]{URL }%
\providecommand \Eprint [0]{\href }%
\providecommand \doibase [0]{http://dx.doi.org/}%
\providecommand \selectlanguage [0]{\@gobble}%
\providecommand \bibinfo  [0]{\@secondoftwo}%
\providecommand \bibfield  [0]{\@secondoftwo}%
\providecommand \translation [1]{[#1]}%
\providecommand \BibitemOpen [0]{}%
\providecommand \bibitemStop [0]{}%
\providecommand \bibitemNoStop [0]{.\EOS\space}%
\providecommand \EOS [0]{\spacefactor3000\relax}%
\providecommand \BibitemShut  [1]{\csname bibitem#1\endcsname}%
\let\auto@bib@innerbib\@empty
\bibitem [{\citenamefont {Crisp}(1970)}]{Crisp1970}%
  \BibitemOpen
  \bibfield  {author} {\bibinfo {author} {\bibfnamefont {M.~D.}\ \bibnamefont
  {Crisp}},\ }\href {\doibase 10.1103/physreva.2.2172.2} {\bibfield  {journal}
  {\bibinfo  {journal} {Physical Review A}\ }\textbf {\bibinfo {volume} {2}},\
  \bibinfo {pages} {2172} (\bibinfo {year} {1970})}\BibitemShut {NoStop}%
\bibitem [{\citenamefont {Bohm}(1952)}]{Bohm1952}%
  \BibitemOpen
  \bibfield  {author} {\bibinfo {author} {\bibfnamefont {D.}~\bibnamefont
  {Bohm}},\ }\href {\doibase 10.1103/PhysRev.85.166} {\bibfield  {journal}
  {\bibinfo  {journal} {Phys. Rev.}\ }\textbf {\bibinfo {volume} {85}},\
  \bibinfo {pages} {166} (\bibinfo {year} {1952})}\BibitemShut {NoStop}%
\bibitem [{\citenamefont {Aharonov}\ \emph {et~al.}(1964)\citenamefont
  {Aharonov}, \citenamefont {Bergmann},\ and\ \citenamefont
  {Lebowitz}}]{ABL1964}%
  \BibitemOpen
  \bibfield  {author} {\bibinfo {author} {\bibfnamefont {Y.}~\bibnamefont
  {Aharonov}}, \bibinfo {author} {\bibfnamefont {P.~G.}\ \bibnamefont
  {Bergmann}}, \ and\ \bibinfo {author} {\bibfnamefont {J.~L.}\ \bibnamefont
  {Lebowitz}},\ }\href {\doibase 10.1103/PhysRev.134.B1410} {\bibfield
  {journal} {\bibinfo  {journal} {Phys. Rev.}\ }\textbf {\bibinfo {volume}
  {134}},\ \bibinfo {pages} {B1410} (\bibinfo {year} {1964})}\BibitemShut
  {NoStop}%
\bibitem [{\citenamefont {Wheeler}\ and\ \citenamefont
  {Zurek}(1983)}]{wheeler_zurek_1983}%
  \BibitemOpen
  \bibfield  {author} {\bibinfo {author} {\bibfnamefont {J.~A.}\ \bibnamefont
  {Wheeler}}\ and\ \bibinfo {author} {\bibfnamefont {W.~H.}\ \bibnamefont
  {Zurek}},\ }\href@noop {} {\emph {\bibinfo {title} {Quantum Theory and
  Measurement}}}\ (\bibinfo  {publisher} {Princeton University Press},\
  \bibinfo {year} {1983})\BibitemShut {NoStop}%
\bibitem [{\citenamefont {Griffiths}(1984)}]{Griffiths1984}%
  \BibitemOpen
  \bibfield  {author} {\bibinfo {author} {\bibfnamefont {R.~B.}\ \bibnamefont
  {Griffiths}},\ }\href {\doibase 10.1007/BF01015734} {\bibfield  {journal}
  {\bibinfo  {journal} {Journal of Statistical Physics}\ }\textbf {\bibinfo
  {volume} {36}},\ \bibinfo {pages} {219} (\bibinfo {year} {1984})}\BibitemShut
  {NoStop}%
\bibitem [{\citenamefont {Aharonov}\ \emph {et~al.}(1988)\citenamefont
  {Aharonov}, \citenamefont {Albert},\ and\ \citenamefont {Vaidman}}]{AAV1988}%
  \BibitemOpen
  \bibfield  {author} {\bibinfo {author} {\bibfnamefont {Y.}~\bibnamefont
  {Aharonov}}, \bibinfo {author} {\bibfnamefont {D.~Z.}\ \bibnamefont
  {Albert}}, \ and\ \bibinfo {author} {\bibfnamefont {L.}~\bibnamefont
  {Vaidman}},\ }\href {\doibase 10.1103/PhysRevLett.60.1351} {\bibfield
  {journal} {\bibinfo  {journal} {Phys. Rev. Lett.}\ }\textbf {\bibinfo
  {volume} {60}},\ \bibinfo {pages} {1351} (\bibinfo {year}
  {1988})}\BibitemShut {NoStop}%
\bibitem [{\citenamefont {Leggett}(1989)}]{Leggett1989}%
  \BibitemOpen
  \bibfield  {author} {\bibinfo {author} {\bibfnamefont {A.~J.}\ \bibnamefont
  {Leggett}},\ }\href {\doibase 10.1103/PhysRevLett.62.2325} {\bibfield
  {journal} {\bibinfo  {journal} {Phys. Rev. Lett.}\ }\textbf {\bibinfo
  {volume} {62}},\ \bibinfo {pages} {2325} (\bibinfo {year}
  {1989})}\BibitemShut {NoStop}%
\bibitem [{\citenamefont {Gell-Mann}\ and\ \citenamefont
  {Hartle}(2018)}]{gellmann2018quantum}%
  \BibitemOpen
  \bibfield  {author} {\bibinfo {author} {\bibfnamefont {M.}~\bibnamefont
  {Gell-Mann}}\ and\ \bibinfo {author} {\bibfnamefont {J.~B.}\ \bibnamefont
  {Hartle}},\ }\href@noop {} {\enquote {\bibinfo {title} {Quantum mechanics in
  the light of quantum cosmology},}\ } (\bibinfo {year} {2018}),\ \Eprint
  {http://arxiv.org/abs/1803.04605} {arXiv:1803.04605 [gr-qc]} \BibitemShut
  {NoStop}%
\bibitem [{\citenamefont {Mir}\ \emph {et~al.}(2007)\citenamefont {Mir},
  \citenamefont {Lundeen}, \citenamefont {Mitchell}, \citenamefont {Steinberg},
  \citenamefont {Garretson},\ and\ \citenamefont {Wiseman}}]{Mir_2007}%
  \BibitemOpen
  \bibfield  {author} {\bibinfo {author} {\bibfnamefont {R.}~\bibnamefont
  {Mir}}, \bibinfo {author} {\bibfnamefont {J.~S.}\ \bibnamefont {Lundeen}},
  \bibinfo {author} {\bibfnamefont {M.~W.}\ \bibnamefont {Mitchell}}, \bibinfo
  {author} {\bibfnamefont {A.~M.}\ \bibnamefont {Steinberg}}, \bibinfo {author}
  {\bibfnamefont {J.~L.}\ \bibnamefont {Garretson}}, \ and\ \bibinfo {author}
  {\bibfnamefont {H.~M.}\ \bibnamefont {Wiseman}},\ }\href {\doibase
  10.1088/1367-2630/9/8/287} {\bibfield  {journal} {\bibinfo  {journal} {New
  Journal of Physics}\ }\textbf {\bibinfo {volume} {9}},\ \bibinfo {pages}
  {287} (\bibinfo {year} {2007})}\BibitemShut {NoStop}%
\bibitem [{\citenamefont {Kocsis}\ \emph {et~al.}(2011)\citenamefont {Kocsis},
  \citenamefont {Braverman}, \citenamefont {Ravets}, \citenamefont {Stevens},
  \citenamefont {Mirin}, \citenamefont {Shalm},\ and\ \citenamefont
  {Steinberg}}]{Kocsis1170}%
  \BibitemOpen
  \bibfield  {author} {\bibinfo {author} {\bibfnamefont {S.}~\bibnamefont
  {Kocsis}}, \bibinfo {author} {\bibfnamefont {B.}~\bibnamefont {Braverman}},
  \bibinfo {author} {\bibfnamefont {S.}~\bibnamefont {Ravets}}, \bibinfo
  {author} {\bibfnamefont {M.~J.}\ \bibnamefont {Stevens}}, \bibinfo {author}
  {\bibfnamefont {R.~P.}\ \bibnamefont {Mirin}}, \bibinfo {author}
  {\bibfnamefont {L.~K.}\ \bibnamefont {Shalm}}, \ and\ \bibinfo {author}
  {\bibfnamefont {A.~M.}\ \bibnamefont {Steinberg}},\ }\href {\doibase
  10.1126/science.1202218} {\bibfield  {journal} {\bibinfo  {journal}
  {Science}\ }\textbf {\bibinfo {volume} {332}},\ \bibinfo {pages} {1170}
  (\bibinfo {year} {2011})},\ \Eprint
  {http://arxiv.org/abs/https://science.sciencemag.org/content/332/6034/1170.full.pdf}
  {https://science.sciencemag.org/content/332/6034/1170.full.pdf} \BibitemShut
  {NoStop}%
\bibitem [{\citenamefont {Lundeen}\ and\ \citenamefont
  {Steinberg}(2009)}]{Lundeen2009}%
  \BibitemOpen
  \bibfield  {author} {\bibinfo {author} {\bibfnamefont {J.~S.}\ \bibnamefont
  {Lundeen}}\ and\ \bibinfo {author} {\bibfnamefont {A.~M.}\ \bibnamefont
  {Steinberg}},\ }\href {\doibase 10.1103/PhysRevLett.102.020404} {\bibfield
  {journal} {\bibinfo  {journal} {Phys. Rev. Lett.}\ }\textbf {\bibinfo
  {volume} {102}},\ \bibinfo {pages} {020404} (\bibinfo {year}
  {2009})}\BibitemShut {NoStop}%
\bibitem [{\citenamefont {Danan}\ \emph {et~al.}(2013)\citenamefont {Danan},
  \citenamefont {Farfurnik}, \citenamefont {Bar-Ad},\ and\ \citenamefont
  {Vaidman}}]{Vaidman2013}%
  \BibitemOpen
  \bibfield  {author} {\bibinfo {author} {\bibfnamefont {A.}~\bibnamefont
  {Danan}}, \bibinfo {author} {\bibfnamefont {D.}~\bibnamefont {Farfurnik}},
  \bibinfo {author} {\bibfnamefont {S.}~\bibnamefont {Bar-Ad}}, \ and\ \bibinfo
  {author} {\bibfnamefont {L.}~\bibnamefont {Vaidman}},\ }\href {\doibase
  10.1103/PhysRevLett.111.240402} {\bibfield  {journal} {\bibinfo  {journal}
  {Phys. Rev. Lett.}\ }\textbf {\bibinfo {volume} {111}},\ \bibinfo {pages}
  {240402} (\bibinfo {year} {2013})}\BibitemShut {NoStop}%
\bibitem [{\citenamefont {Gammelmark}\ \emph {et~al.}(2014)\citenamefont
  {Gammelmark}, \citenamefont {M\o{}lmer}, \citenamefont {Alt}, \citenamefont
  {Kampschulte},\ and\ \citenamefont {Meschede}}]{Gammelmark2014}%
  \BibitemOpen
  \bibfield  {author} {\bibinfo {author} {\bibfnamefont {S.}~\bibnamefont
  {Gammelmark}}, \bibinfo {author} {\bibfnamefont {K.}~\bibnamefont
  {M\o{}lmer}}, \bibinfo {author} {\bibfnamefont {W.}~\bibnamefont {Alt}},
  \bibinfo {author} {\bibfnamefont {T.}~\bibnamefont {Kampschulte}}, \ and\
  \bibinfo {author} {\bibfnamefont {D.}~\bibnamefont {Meschede}},\ }\href
  {\doibase 10.1103/PhysRevA.89.043839} {\bibfield  {journal} {\bibinfo
  {journal} {Phys. Rev. A}\ }\textbf {\bibinfo {volume} {89}},\ \bibinfo
  {pages} {043839} (\bibinfo {year} {2014})}\BibitemShut {NoStop}%
\bibitem [{\citenamefont {Guevara}\ and\ \citenamefont
  {Wiseman}(2018)}]{Guevara2018}%
  \BibitemOpen
  \bibfield  {author} {\bibinfo {author} {\bibfnamefont {I.}~\bibnamefont
  {Guevara}}\ and\ \bibinfo {author} {\bibfnamefont {H.}~\bibnamefont
  {Wiseman}},\ }\href@noop {} {\emph {\bibinfo {title} {{Quantum State
  Smoothing}}}},\ \bibinfo {type} {Tech. Rep.}\ (\bibinfo {year} {2018})\
  \Eprint {http://arxiv.org/abs/1503.02799v3} {arXiv:1503.02799v3} \BibitemShut
  {NoStop}%
\bibitem [{\citenamefont {Tsang}(2009)}]{Tsang2009}%
  \BibitemOpen
  \bibfield  {author} {\bibinfo {author} {\bibfnamefont {M.}~\bibnamefont
  {Tsang}},\ }\href@noop {} {\emph {\bibinfo {title} {{Time-Symmetric Quantum
  Theory of Smoothing}}}},\ \bibinfo {type} {Tech. Rep.}\ (\bibinfo {year}
  {2009})\ \Eprint {http://arxiv.org/abs/0904.1969v2} {arXiv:0904.1969v2}
  \BibitemShut {NoStop}%
\bibitem [{\citenamefont {Rybarczyk}\ \emph {et~al.}(2015)\citenamefont
  {Rybarczyk}, \citenamefont {Peaudecerf}, \citenamefont {Penasa},
  \citenamefont {Gerlich}, \citenamefont {Julsgaard}, \citenamefont
  {M\o{}lmer}, \citenamefont {Gleyzes}, \citenamefont {Brune}, \citenamefont
  {Raimond}, \citenamefont {Haroche},\ and\ \citenamefont
  {Dotsenko}}]{Rybarczyk2015}%
  \BibitemOpen
  \bibfield  {author} {\bibinfo {author} {\bibfnamefont {T.}~\bibnamefont
  {Rybarczyk}}, \bibinfo {author} {\bibfnamefont {B.}~\bibnamefont
  {Peaudecerf}}, \bibinfo {author} {\bibfnamefont {M.}~\bibnamefont {Penasa}},
  \bibinfo {author} {\bibfnamefont {S.}~\bibnamefont {Gerlich}}, \bibinfo
  {author} {\bibfnamefont {B.}~\bibnamefont {Julsgaard}}, \bibinfo {author}
  {\bibfnamefont {K.}~\bibnamefont {M\o{}lmer}}, \bibinfo {author}
  {\bibfnamefont {S.}~\bibnamefont {Gleyzes}}, \bibinfo {author} {\bibfnamefont
  {M.}~\bibnamefont {Brune}}, \bibinfo {author} {\bibfnamefont {J.~M.}\
  \bibnamefont {Raimond}}, \bibinfo {author} {\bibfnamefont {S.}~\bibnamefont
  {Haroche}}, \ and\ \bibinfo {author} {\bibfnamefont {I.}~\bibnamefont
  {Dotsenko}},\ }\href {\doibase 10.1103/PhysRevA.91.062116} {\bibfield
  {journal} {\bibinfo  {journal} {Phys. Rev. A}\ }\textbf {\bibinfo {volume}
  {91}},\ \bibinfo {pages} {062116} (\bibinfo {year} {2015})}\BibitemShut
  {NoStop}%
\bibitem [{\citenamefont {Hacohen-Gourgy}\ \emph {et~al.}(2016)\citenamefont
  {Hacohen-Gourgy}, \citenamefont {Martin}, \citenamefont {Flurin},
  \citenamefont {Ramasesh}, \citenamefont {Whaley},\ and\ \citenamefont
  {Siddiqi}}]{HacohenGourgy2016}%
  \BibitemOpen
  \bibfield  {author} {\bibinfo {author} {\bibfnamefont {S.}~\bibnamefont
  {Hacohen-Gourgy}}, \bibinfo {author} {\bibfnamefont {L.~S.}\ \bibnamefont
  {Martin}}, \bibinfo {author} {\bibfnamefont {E.}~\bibnamefont {Flurin}},
  \bibinfo {author} {\bibfnamefont {V.~V.}\ \bibnamefont {Ramasesh}}, \bibinfo
  {author} {\bibfnamefont {K.~B.}\ \bibnamefont {Whaley}}, \ and\ \bibinfo
  {author} {\bibfnamefont {I.}~\bibnamefont {Siddiqi}},\ }\href {\doibase
  10.1038/nature19762} {\bibfield  {journal} {\bibinfo  {journal} {Nature}\
  }\textbf {\bibinfo {volume} {538}},\ \bibinfo {pages} {491} (\bibinfo {year}
  {2016})}\BibitemShut {NoStop}%
\bibitem [{\citenamefont {Minev}\ \emph {et~al.}(2019)\citenamefont {Minev},
  \citenamefont {Mundhada}, \citenamefont {Shankar}, \citenamefont {Reinhold},
  \citenamefont {Guti{\'e}rrez-J{\'a}uregui}, \citenamefont {Schoelkopf},
  \citenamefont {Mirrahimi}, \citenamefont {Carmichael},\ and\ \citenamefont
  {Devoret}}]{Minev2019}%
  \BibitemOpen
  \bibfield  {author} {\bibinfo {author} {\bibfnamefont {Z.}~\bibnamefont
  {Minev}}, \bibinfo {author} {\bibfnamefont {S.}~\bibnamefont {Mundhada}},
  \bibinfo {author} {\bibfnamefont {S.}~\bibnamefont {Shankar}}, \bibinfo
  {author} {\bibfnamefont {P.}~\bibnamefont {Reinhold}}, \bibinfo {author}
  {\bibfnamefont {R.}~\bibnamefont {Guti{\'e}rrez-J{\'a}uregui}}, \bibinfo
  {author} {\bibfnamefont {R.}~\bibnamefont {Schoelkopf}}, \bibinfo {author}
  {\bibfnamefont {M.}~\bibnamefont {Mirrahimi}}, \bibinfo {author}
  {\bibfnamefont {H.}~\bibnamefont {Carmichael}}, \ and\ \bibinfo {author}
  {\bibfnamefont {M.}~\bibnamefont {Devoret}},\ }\href {\doibase
  10.1038/s41586-019-1287-z} {\bibfield  {journal} {\bibinfo  {journal}
  {Nature}\ }\textbf {\bibinfo {volume} {570}},\ \bibinfo {pages} {200}
  (\bibinfo {year} {2019})}\BibitemShut {NoStop}%
\bibitem [{\citenamefont {Turchette}\ \emph {et~al.}(1995)\citenamefont
  {Turchette}, \citenamefont {Hood}, \citenamefont {Lange}, \citenamefont
  {Mabuchi},\ and\ \citenamefont {Kimble}}]{Kimble1995}%
  \BibitemOpen
  \bibfield  {author} {\bibinfo {author} {\bibfnamefont {Q.~A.}\ \bibnamefont
  {Turchette}}, \bibinfo {author} {\bibfnamefont {C.~J.}\ \bibnamefont {Hood}},
  \bibinfo {author} {\bibfnamefont {W.}~\bibnamefont {Lange}}, \bibinfo
  {author} {\bibfnamefont {H.}~\bibnamefont {Mabuchi}}, \ and\ \bibinfo
  {author} {\bibfnamefont {H.~J.}\ \bibnamefont {Kimble}},\ }\href {\doibase
  10.1103/PhysRevLett.75.4710} {\bibfield  {journal} {\bibinfo  {journal}
  {Phys. Rev. Lett.}\ }\textbf {\bibinfo {volume} {75}},\ \bibinfo {pages}
  {4710} (\bibinfo {year} {1995})}\BibitemShut {NoStop}%
\bibitem [{\citenamefont {Schmidt}\ and\ \citenamefont
  {Imamoglu}(1996)}]{Schmidt96}%
  \BibitemOpen
  \bibfield  {author} {\bibinfo {author} {\bibfnamefont {H.}~\bibnamefont
  {Schmidt}}\ and\ \bibinfo {author} {\bibfnamefont {A.}~\bibnamefont
  {Imamoglu}},\ }\href {\doibase 10.1364/OL.21.001936} {\bibfield  {journal}
  {\bibinfo  {journal} {Opt. Lett.}\ }\textbf {\bibinfo {volume} {21}},\
  \bibinfo {pages} {1936} (\bibinfo {year} {1996})}\BibitemShut {NoStop}%
\bibitem [{\citenamefont {Venkataraman}\ \emph {et~al.}(2013)\citenamefont
  {Venkataraman}, \citenamefont {Saha},\ and\ \citenamefont
  {Gaeta}}]{Venkataraman2013}%
  \BibitemOpen
  \bibfield  {author} {\bibinfo {author} {\bibfnamefont {V.}~\bibnamefont
  {Venkataraman}}, \bibinfo {author} {\bibfnamefont {K.}~\bibnamefont {Saha}},
  \ and\ \bibinfo {author} {\bibfnamefont {A.~L.}\ \bibnamefont {Gaeta}},\
  }\href {\doibase 10.1038/nphoton.2012.283} {\bibfield  {journal} {\bibinfo
  {journal} {Nature Photonics}\ }\textbf {\bibinfo {volume} {7}},\ \bibinfo
  {pages} {138} (\bibinfo {year} {2013})}\BibitemShut {NoStop}%
\bibitem [{\citenamefont {Feizpour}\ \emph {et~al.}(2015)\citenamefont
  {Feizpour}, \citenamefont {Hallaji}, \citenamefont {Dmochowski},\ and\
  \citenamefont {Steinberg}}]{Feizpour2015a}%
  \BibitemOpen
  \bibfield  {author} {\bibinfo {author} {\bibfnamefont {A.}~\bibnamefont
  {Feizpour}}, \bibinfo {author} {\bibfnamefont {M.}~\bibnamefont {Hallaji}},
  \bibinfo {author} {\bibfnamefont {G.}~\bibnamefont {Dmochowski}}, \ and\
  \bibinfo {author} {\bibfnamefont {A.~M.}\ \bibnamefont {Steinberg}},\ }\href
  {\doibase 10.1038/nphys3433} {\bibfield  {journal} {\bibinfo  {journal}
  {Nature Physics}\ }\textbf {\bibinfo {volume} {11}},\ \bibinfo {pages} {905}
  (\bibinfo {year} {2015})}\BibitemShut {NoStop}%
\bibitem [{\citenamefont {Tiarks}\ \emph {et~al.}(2016)\citenamefont {Tiarks},
  \citenamefont {Schmidt}, \citenamefont {Rempe},\ and\ \citenamefont
  {D{\"u}rr}}]{Tiarkse1600036}%
  \BibitemOpen
  \bibfield  {author} {\bibinfo {author} {\bibfnamefont {D.}~\bibnamefont
  {Tiarks}}, \bibinfo {author} {\bibfnamefont {S.}~\bibnamefont {Schmidt}},
  \bibinfo {author} {\bibfnamefont {G.}~\bibnamefont {Rempe}}, \ and\ \bibinfo
  {author} {\bibfnamefont {S.}~\bibnamefont {D{\"u}rr}},\ }\href {\doibase
  10.1126/sciadv.1600036} {\bibfield  {journal} {\bibinfo  {journal} {Science
  Advances}\ }\textbf {\bibinfo {volume} {2}} (\bibinfo {year} {2016}),\
  10.1126/sciadv.1600036},\ \Eprint
  {http://arxiv.org/abs/https://advances.sciencemag.org/content/2/4/e1600036.full.pdf}
  {https://advances.sciencemag.org/content/2/4/e1600036.full.pdf} \BibitemShut
  {NoStop}%
\bibitem [{\citenamefont {Sinclair}\ \emph {et~al.}(2019)\citenamefont
  {Sinclair}, \citenamefont {Angulo}, \citenamefont {Lupu-Gladstein},
  \citenamefont {Bonsma-Fisher},\ and\ \citenamefont
  {Steinberg}}]{Sinclair2019}%
  \BibitemOpen
  \bibfield  {author} {\bibinfo {author} {\bibfnamefont {J.}~\bibnamefont
  {Sinclair}}, \bibinfo {author} {\bibfnamefont {D.}~\bibnamefont {Angulo}},
  \bibinfo {author} {\bibfnamefont {N.}~\bibnamefont {Lupu-Gladstein}},
  \bibinfo {author} {\bibfnamefont {K.}~\bibnamefont {Bonsma-Fisher}}, \ and\
  \bibinfo {author} {\bibfnamefont {A.~M.}\ \bibnamefont {Steinberg}},\ }\href
  {\doibase 10.1103/PhysRevResearch.1.033193} {\bibfield  {journal} {\bibinfo
  {journal} {Phys. Rev. Research}\ }\textbf {\bibinfo {volume} {1}},\ \bibinfo
  {pages} {033193} (\bibinfo {year} {2019})}\BibitemShut {NoStop}%
\bibitem [{Note1()}]{Note1}%
  \BibitemOpen
  \bibinfo {note} {This is clearly closely related to the weak measurements of
  Aharonov, Albert and Vaidman, as it involves a pre- and post-selection, and a
  weak coupling to a probe system; for other examples of pre- and post-selected
  systems including from Cavity QED see \cite {Foster2000,Smith2002,
  Wiseman2002, Duan2020})}\BibitemShut {NoStop}%
\bibitem [{\citenamefont {Thompson}\ \emph {et~al.}(2006)\citenamefont
  {Thompson}, \citenamefont {Simon}, \citenamefont {Loh},\ and\ \citenamefont
  {Vuleti{\'c}}}]{Thompson74}%
  \BibitemOpen
  \bibfield  {author} {\bibinfo {author} {\bibfnamefont {J.~K.}\ \bibnamefont
  {Thompson}}, \bibinfo {author} {\bibfnamefont {J.}~\bibnamefont {Simon}},
  \bibinfo {author} {\bibfnamefont {H.}~\bibnamefont {Loh}}, \ and\ \bibinfo
  {author} {\bibfnamefont {V.}~\bibnamefont {Vuleti{\'c}}},\ }\href {\doibase
  10.1126/science.1127676} {\bibfield  {journal} {\bibinfo  {journal}
  {Science}\ }\textbf {\bibinfo {volume} {313}},\ \bibinfo {pages} {74}
  (\bibinfo {year} {2006})}\BibitemShut {NoStop}%
\bibitem [{\citenamefont {Xing}(2013)}]{Xing2013}%
  \BibitemOpen
  \bibfield  {author} {\bibinfo {author} {\bibfnamefont {X.}~\bibnamefont
  {Xing}},\ }\href@noop {} {\enquote {\bibinfo {title} {{A Quantum Light Source
  for Light-Matter Interaction}},}\ } (\bibinfo {year} {2013})\BibitemShut
  {NoStop}%
\bibitem [{\citenamefont {Loredo}\ \emph {et~al.}(2016)\citenamefont {Loredo},
  \citenamefont {Zakaria}, \citenamefont {Somaschi}, \citenamefont {Anton},
  \citenamefont {de~Santis}, \citenamefont {Giesz}, \citenamefont {Grange},
  \citenamefont {Broome}, \citenamefont {Gazzano}, \citenamefont {Coppola},
  \citenamefont {Sagnes}, \citenamefont {Lemaitre}, \citenamefont {Auffeves},
  \citenamefont {Senellart}, \citenamefont {Almeida},\ and\ \citenamefont
  {White}}]{Loredo2016}%
  \BibitemOpen
  \bibfield  {author} {\bibinfo {author} {\bibfnamefont {J.~C.}\ \bibnamefont
  {Loredo}}, \bibinfo {author} {\bibfnamefont {N.~A.}\ \bibnamefont {Zakaria}},
  \bibinfo {author} {\bibfnamefont {N.}~\bibnamefont {Somaschi}}, \bibinfo
  {author} {\bibfnamefont {C.}~\bibnamefont {Anton}}, \bibinfo {author}
  {\bibfnamefont {L.}~\bibnamefont {de~Santis}}, \bibinfo {author}
  {\bibfnamefont {V.}~\bibnamefont {Giesz}}, \bibinfo {author} {\bibfnamefont
  {T.}~\bibnamefont {Grange}}, \bibinfo {author} {\bibfnamefont {M.~A.}\
  \bibnamefont {Broome}}, \bibinfo {author} {\bibfnamefont {O.}~\bibnamefont
  {Gazzano}}, \bibinfo {author} {\bibfnamefont {G.}~\bibnamefont {Coppola}},
  \bibinfo {author} {\bibfnamefont {I.}~\bibnamefont {Sagnes}}, \bibinfo
  {author} {\bibfnamefont {A.}~\bibnamefont {Lemaitre}}, \bibinfo {author}
  {\bibfnamefont {A.}~\bibnamefont {Auffeves}}, \bibinfo {author}
  {\bibfnamefont {P.}~\bibnamefont {Senellart}}, \bibinfo {author}
  {\bibfnamefont {M.~P.}\ \bibnamefont {Almeida}}, \ and\ \bibinfo {author}
  {\bibfnamefont {A.~G.}\ \bibnamefont {White}},\ }\href@noop {} {\bibfield
  {journal} {\bibinfo  {journal} {Optica}\ }\textbf {\bibinfo {volume} {3}},\
  \bibinfo {pages} {433} (\bibinfo {year} {2016})}\BibitemShut {NoStop}%
\bibitem [{\citenamefont {Wang}\ \emph {et~al.}(2016)\citenamefont {Wang},
  \citenamefont {Chen}, \citenamefont {Li}, \citenamefont {Huang},
  \citenamefont {Liu}, \citenamefont {Chen}, \citenamefont {Luo}, \citenamefont
  {Su}, \citenamefont {Wu}, \citenamefont {Li}, \citenamefont {Lu},
  \citenamefont {Hu}, \citenamefont {Jiang}, \citenamefont {Peng},
  \citenamefont {Li}, \citenamefont {Liu}, \citenamefont {Chen}, \citenamefont
  {Lu},\ and\ \citenamefont {Pan}}]{JianWeiPan2016}%
  \BibitemOpen
  \bibfield  {author} {\bibinfo {author} {\bibfnamefont {X.-L.}\ \bibnamefont
  {Wang}}, \bibinfo {author} {\bibfnamefont {L.-K.}\ \bibnamefont {Chen}},
  \bibinfo {author} {\bibfnamefont {W.}~\bibnamefont {Li}}, \bibinfo {author}
  {\bibfnamefont {H.-L.}\ \bibnamefont {Huang}}, \bibinfo {author}
  {\bibfnamefont {C.}~\bibnamefont {Liu}}, \bibinfo {author} {\bibfnamefont
  {C.}~\bibnamefont {Chen}}, \bibinfo {author} {\bibfnamefont {Y.-H.}\
  \bibnamefont {Luo}}, \bibinfo {author} {\bibfnamefont {Z.-E.}\ \bibnamefont
  {Su}}, \bibinfo {author} {\bibfnamefont {D.}~\bibnamefont {Wu}}, \bibinfo
  {author} {\bibfnamefont {Z.-D.}\ \bibnamefont {Li}}, \bibinfo {author}
  {\bibfnamefont {H.}~\bibnamefont {Lu}}, \bibinfo {author} {\bibfnamefont
  {Y.}~\bibnamefont {Hu}}, \bibinfo {author} {\bibfnamefont {X.}~\bibnamefont
  {Jiang}}, \bibinfo {author} {\bibfnamefont {C.-Z.}\ \bibnamefont {Peng}},
  \bibinfo {author} {\bibfnamefont {L.}~\bibnamefont {Li}}, \bibinfo {author}
  {\bibfnamefont {N.-L.}\ \bibnamefont {Liu}}, \bibinfo {author} {\bibfnamefont
  {Y.-A.}\ \bibnamefont {Chen}}, \bibinfo {author} {\bibfnamefont {C.-Y.}\
  \bibnamefont {Lu}}, \ and\ \bibinfo {author} {\bibfnamefont {J.-W.}\
  \bibnamefont {Pan}},\ }\href {\doibase 10.1103/PhysRevLett.117.210502}
  {\bibfield  {journal} {\bibinfo  {journal} {Phys. Rev. Lett.}\ }\textbf
  {\bibinfo {volume} {117}},\ \bibinfo {pages} {210502} (\bibinfo {year}
  {2016})}\BibitemShut {NoStop}%
\bibitem [{\citenamefont {Ornelas-Huerta}\ \emph {et~al.}(2020)\citenamefont
  {Ornelas-Huerta}, \citenamefont {Craddock}, \citenamefont {Goldschmidt},
  \citenamefont {Hachtel}, \citenamefont {Wang}, \citenamefont {Bienias},
  \citenamefont {Gorshkov}, \citenamefont {Rolston},\ and\ \citenamefont
  {Porto}}]{ornelashuerta2020}%
  \BibitemOpen
  \bibfield  {author} {\bibinfo {author} {\bibfnamefont {D.~P.}\ \bibnamefont
  {Ornelas-Huerta}}, \bibinfo {author} {\bibfnamefont {A.~N.}\ \bibnamefont
  {Craddock}}, \bibinfo {author} {\bibfnamefont {E.~A.}\ \bibnamefont
  {Goldschmidt}}, \bibinfo {author} {\bibfnamefont {A.~J.}\ \bibnamefont
  {Hachtel}}, \bibinfo {author} {\bibfnamefont {Y.}~\bibnamefont {Wang}},
  \bibinfo {author} {\bibfnamefont {P.}~\bibnamefont {Bienias}}, \bibinfo
  {author} {\bibfnamefont {A.~V.}\ \bibnamefont {Gorshkov}}, \bibinfo {author}
  {\bibfnamefont {S.~L.}\ \bibnamefont {Rolston}}, \ and\ \bibinfo {author}
  {\bibfnamefont {J.~V.}\ \bibnamefont {Porto}},\ }\href {\doibase
  10.1364/OPTICA.391485} {\bibfield  {journal} {\bibinfo  {journal} {Optica}\
  }\textbf {\bibinfo {volume} {7}},\ \bibinfo {pages} {813} (\bibinfo {year}
  {2020})}\BibitemShut {NoStop}%
\bibitem [{\citenamefont {Hallaji}\ \emph {et~al.}(2017)\citenamefont
  {Hallaji}, \citenamefont {Feizpour}, \citenamefont {Dmochowski},
  \citenamefont {Sinclair},\ and\ \citenamefont {Steinberg}}]{Hallaji2017}%
  \BibitemOpen
  \bibfield  {author} {\bibinfo {author} {\bibfnamefont {M.}~\bibnamefont
  {Hallaji}}, \bibinfo {author} {\bibfnamefont {A.}~\bibnamefont {Feizpour}},
  \bibinfo {author} {\bibfnamefont {G.}~\bibnamefont {Dmochowski}}, \bibinfo
  {author} {\bibfnamefont {J.}~\bibnamefont {Sinclair}}, \ and\ \bibinfo
  {author} {\bibfnamefont {A.}~\bibnamefont {Steinberg}},\ }\href {\doibase
  10.1038/nphys4040} {\bibfield  {journal} {\bibinfo  {journal} {Nature
  Physics}\ }\textbf {\bibinfo {volume} {13}},\ \bibinfo {pages} {540}
  (\bibinfo {year} {2017})}\BibitemShut {NoStop}%
\bibitem [{\citenamefont {McCall}\ and\ \citenamefont
  {Hahn}(1967)}]{McCall1967a}%
  \BibitemOpen
  \bibfield  {author} {\bibinfo {author} {\bibfnamefont {S.~L.}\ \bibnamefont
  {McCall}}\ and\ \bibinfo {author} {\bibfnamefont {E.~L.}\ \bibnamefont
  {Hahn}},\ }\href {\doibase 10.1103/PhysRevLett.18.908} {\bibfield  {journal}
  {\bibinfo  {journal} {Phys. Rev. Lett.}\ }\textbf {\bibinfo {volume} {18}},\
  \bibinfo {pages} {908} (\bibinfo {year} {1967})}\BibitemShut {NoStop}%
\bibitem [{\citenamefont {McCall}\ and\ \citenamefont
  {Hahn}(1969)}]{McCall1967b}%
  \BibitemOpen
  \bibfield  {author} {\bibinfo {author} {\bibfnamefont {S.~L.}\ \bibnamefont
  {McCall}}\ and\ \bibinfo {author} {\bibfnamefont {E.~L.}\ \bibnamefont
  {Hahn}},\ }\href {\doibase 10.1103/PhysRev.183.457} {\bibfield  {journal}
  {\bibinfo  {journal} {Phys. Rev.}\ }\textbf {\bibinfo {volume} {183}},\
  \bibinfo {pages} {457} (\bibinfo {year} {1969})}\BibitemShut {NoStop}%
\bibitem [{\citenamefont {Allen}\ and\ \citenamefont
  {Eberly}(1975)}]{AllenEberlyMonograph}%
  \BibitemOpen
  \bibfield  {author} {\bibinfo {author} {\bibfnamefont {L.}~\bibnamefont
  {Allen}}\ and\ \bibinfo {author} {\bibfnamefont {J.}~\bibnamefont {Eberly}},\
  }\href@noop {} {\emph {\bibinfo {title} {Optical Resonance and Two-Level
  Atoms}}}\ (\bibinfo  {publisher} {John Wiley and Sons, Inc},\ \bibinfo {year}
  {1975})\BibitemShut {NoStop}%
\bibitem [{\citenamefont {Costanzo}\ \emph {et~al.}(2016)\citenamefont
  {Costanzo}, \citenamefont {Coelho}, \citenamefont {Pellegrino}, \citenamefont
  {Mendes}, \citenamefont {Acioli}, \citenamefont {Cassemiro}, \citenamefont
  {Felinto}, \citenamefont {Zavatta},\ and\ \citenamefont
  {Bellini}}]{Costanzo2016}%
  \BibitemOpen
  \bibfield  {author} {\bibinfo {author} {\bibfnamefont {L.~S.}\ \bibnamefont
  {Costanzo}}, \bibinfo {author} {\bibfnamefont {A.~S.}\ \bibnamefont
  {Coelho}}, \bibinfo {author} {\bibfnamefont {D.}~\bibnamefont {Pellegrino}},
  \bibinfo {author} {\bibfnamefont {M.~S.}\ \bibnamefont {Mendes}}, \bibinfo
  {author} {\bibfnamefont {L.}~\bibnamefont {Acioli}}, \bibinfo {author}
  {\bibfnamefont {K.~N.}\ \bibnamefont {Cassemiro}}, \bibinfo {author}
  {\bibfnamefont {D.}~\bibnamefont {Felinto}}, \bibinfo {author} {\bibfnamefont
  {A.}~\bibnamefont {Zavatta}}, \ and\ \bibinfo {author} {\bibfnamefont
  {M.}~\bibnamefont {Bellini}},\ }\href {\doibase
  10.1103/PhysRevLett.116.023602} {\bibfield  {journal} {\bibinfo  {journal}
  {Phys. Rev. Lett.}\ }\textbf {\bibinfo {volume} {116}},\ \bibinfo {pages}
  {023602} (\bibinfo {year} {2016})}\BibitemShut {NoStop}%
\bibitem [{\citenamefont {Foster}\ \emph {et~al.}(2000)\citenamefont {Foster},
  \citenamefont {Orozco}, \citenamefont {Castro-Beltran},\ and\ \citenamefont
  {Carmichael}}]{Foster2000}%
  \BibitemOpen
  \bibfield  {author} {\bibinfo {author} {\bibfnamefont {G.~T.}\ \bibnamefont
  {Foster}}, \bibinfo {author} {\bibfnamefont {L.~A.}\ \bibnamefont {Orozco}},
  \bibinfo {author} {\bibfnamefont {H.~M.}\ \bibnamefont {Castro-Beltran}}, \
  and\ \bibinfo {author} {\bibfnamefont {H.~J.}\ \bibnamefont {Carmichael}},\
  }\href {\doibase 10.1103/PhysRevLett.85.3149} {\bibfield  {journal} {\bibinfo
   {journal} {Phys. Rev. Lett.}\ }\textbf {\bibinfo {volume} {85}},\ \bibinfo
  {pages} {3149} (\bibinfo {year} {2000})}\BibitemShut {NoStop}%
\bibitem [{\citenamefont {Smith}\ \emph {et~al.}(2002)\citenamefont {Smith},
  \citenamefont {Reiner}, \citenamefont {Orozco}, \citenamefont {Kuhr},\ and\
  \citenamefont {Wiseman}}]{Smith2002}%
  \BibitemOpen
  \bibfield  {author} {\bibinfo {author} {\bibfnamefont {W.~P.}\ \bibnamefont
  {Smith}}, \bibinfo {author} {\bibfnamefont {J.~E.}\ \bibnamefont {Reiner}},
  \bibinfo {author} {\bibfnamefont {L.~A.}\ \bibnamefont {Orozco}}, \bibinfo
  {author} {\bibfnamefont {S.}~\bibnamefont {Kuhr}}, \ and\ \bibinfo {author}
  {\bibfnamefont {H.~M.}\ \bibnamefont {Wiseman}},\ }\href {\doibase
  10.1103/PhysRevLett.89.133601} {\bibfield  {journal} {\bibinfo  {journal}
  {Phys. Rev. Lett.}\ }\textbf {\bibinfo {volume} {89}},\ \bibinfo {pages}
  {133601} (\bibinfo {year} {2002})}\BibitemShut {NoStop}%
\bibitem [{\citenamefont {Wiseman}(2002)}]{Wiseman2002}%
  \BibitemOpen
  \bibfield  {author} {\bibinfo {author} {\bibfnamefont {H.~M.}\ \bibnamefont
  {Wiseman}},\ }\href {\doibase 10.1103/PhysRevA.65.032111} {\bibfield
  {journal} {\bibinfo  {journal} {Phys. Rev. A}\ }\textbf {\bibinfo {volume}
  {65}},\ \bibinfo {pages} {032111} (\bibinfo {year} {2002})}\BibitemShut
  {NoStop}%
\bibitem [{\citenamefont {Duan}\ \emph {et~al.}(2020)\citenamefont {Duan},
  \citenamefont {Hosseini}, \citenamefont {Beck},\ and\ \citenamefont
  {Vuleti\ifmmode~\acute{c}\else \'{c}\fi{}}}]{Duan2020}%
  \BibitemOpen
  \bibfield  {author} {\bibinfo {author} {\bibfnamefont {Y.}~\bibnamefont
  {Duan}}, \bibinfo {author} {\bibfnamefont {M.}~\bibnamefont {Hosseini}},
  \bibinfo {author} {\bibfnamefont {K.~M.}\ \bibnamefont {Beck}}, \ and\
  \bibinfo {author} {\bibfnamefont {V.}~\bibnamefont
  {Vuleti\ifmmode~\acute{c}\else \'{c}\fi{}}},\ }\href {\doibase
  10.1103/PhysRevLett.124.223602} {\bibfield  {journal} {\bibinfo  {journal}
  {Phys. Rev. Lett.}\ }\textbf {\bibinfo {volume} {124}},\ \bibinfo {pages}
  {223602} (\bibinfo {year} {2020})}\BibitemShut {NoStop}%
\end{thebibliography}%

\end{document}